\pdfoutput=1

\documentclass[11pt]{article}

\usepackage[preprint]{acl}
\usepackage{amsmath}
\usepackage{multirow} 

\usepackage{times}
\usepackage{latexsym}

\usepackage[T1]{fontenc}

\usepackage[utf8]{inputenc}

\usepackage{microtype}

\usepackage{inconsolata}

\usepackage{graphicx}

\title{SEARA: An Automated Approach for Obtaining Optimal Retrievers}

\author{
 \textbf{Zou Yuheng},
 \textbf{Wang Yiran},
 \textbf{Tian Yuzhu},
 \textbf{Zhu Min},
 \textbf{Huang Yanhua\textsuperscript{1}}
\\
 \textsuperscript{1}rednote,
\\
 \small{
   \textbf{{loganzou0421@163.com, xingwen@xiaohongshu.com , ytia0933@uni.sydney.edu.au, youer@xiaohongshu.com, teto@xiaohongshu.com}
 }
}
}

\begin{document}
\maketitle
\begin{abstract}
Retrieval-Augmented Generation (RAG) is a core approach for enhancing Large Language Models (LLMs), where the effectiveness of the retriever largely determines the overall response quality of RAG systems. Retrievers encompass a multitude of hyperparameters that significantly impact performance outcomes and demonstrate sensitivity to specific applications. Nevertheless, hyperparameter optimization entails prohibitively high computational expenses. Existing evaluation methods suffer from either prohibitive costs or disconnection from domain-specific scenarios. This paper proposes SEARA (Subset sampling Evaluation for Automatic Retriever Assessment), which addresses evaluation data challenges through subset sampling techniques and achieves robust automated retriever evaluation by minimal retrieval facts extraction and comprehensive retrieval metrics. Based on real user queries, this method enables fully automated retriever evaluation at low cost, thereby obtaining optimal retriever for specific business scenarios. We validate our method across classic RAG applications in rednote, including knowledge-based Q\&A system and retrieval-based travel assistant, successfully obtaining scenario-specific optimal retrievers.
\end{abstract}
\section{Introduction}

The Retrieval-Augmented Generation (RAG) has emerged as one of the most prevalent approaches for AI application development, encompassing both classic RAG scenarios such as knowledge-based Q\&A systems and broader RAG scenarios such as generative retrieval applications~\citep{lewis2020retrieval}. In these systems, the retrieval effectiveness of the retriever largely determines the overall application response quality. However, retriever construction involves multiple strategies including text segmentation, knowledge retrieval, and reranking approaches, with different business scenarios often requiring distinct optimal strategy combinations~\citep{yu2024evaluation}. Determining the optimal retriever construction strategy for specific business scenarios cannot rely solely on prior knowledge.

Experimental search for optimal retriever necessitates a standardized evaluation framework to determine relative performance across specific business scenarios. However, annotating all relevant knowledge for each query to obtain complete ground truth(GT) requires multiple traversals of the entire knowledge base, resulting in prohibitive costs. Meanwhile, Automated RAG evaluation systems like Ragas~\citep{es2024ragas} achieve assessment by synthesizing queries from specified knowledge, which differs from real user queries and fails to guarantee consistency between evaluation results and actual performance.

Therefore, we propose SEARA, an automated approach for obtaining optimal retrievers that efficiently enables fully automated retriever evaluation using real user queries. The core advantages of our method include:
\begin{itemize}
    \item \textbf{Cost-effective evaluation framework}: We address retriever evaluation data challenges through innovative subset sampling techniques, enabling efficient comparison of relative performance across different retrievers. 
    \item \textbf{Universal automated assessment}: We leverage LLMs for intelligent sample discrimination and minimal retrieval fact extraction, accommodating diverse retriever construction strategies and achieving universal automated evaluation across RAG scenarios.
    \item \textbf{Comprehensive evaluation metrics}: We introduce holistic evaluation metrics that balance precision and recall considerations, resolving comparability issues between heterogeneous retrievers and enabling more robust performance assessment.
\end{itemize}

This paper will introduce the details of the SEARA method in the Core Challenges and Solutions section, employ this method in two RAG applications at rednote in the Cases section, and finally present other research work related to this paper in the Related Work section.
\section{Core Challenges and Solutions}
In this section, we introduce three core challenges in retriever evaluation within RAG systems and how our method addresses these challenges to obtain the optimal retriever.

\subsection{Data Acquisition}

\textbf{Core Challenge.}
A retriever ideally achieves perfect recall and precision by retrieving all relevant knowledge without redundancy, but evaluating this performance requires ground truth (GT) annotation of all relevant facts for each query, which is prohibitively expensive to obtain. Manual annotation, while straightforward, is time-consuming and costly. Model-based approaches offer automation through embedding models~\citep{sawarkar2024blended}, rerank models, or LLMs\citep{es2024ragas}. However, embedding and rerank models are limited by their performance capabilities, while LLM-based annotation, though more accurate, results in excessive computational costs.

To address the prohibitive cost of LLM recall, random sampling can be applied to the knowledge base to reduce its size and overall computational cost. For example, randomly sampling a knowledge base $|D| = N$ to $|D'| = N'$, where $N' << N$, achieves final computational cost $M\times N' << M\times N$. 

However, the premise that evaluation effectiveness on the sampled knowledge base aligns with actual evaluation effectiveness requires that the sampled data distribution matches the original knowledge base distribution. Given the complexity of RAG application scenarios, it is difficult to determine an objective, unified sampling method that can eliminate distributional differences. Therefore, evaluation methods based on random sampling lack stable accuracy.

\textbf{Proposed Method: Subset Sampling}. Considering that our core objective is to obtain optimal retrievers for specific business scenarios, and given the difficulty of acquiring complete GT to calculate absolutely accurate precision and recall, if we can construct alternative GT to enable relatively comparable precision and recall calculations, we can still achieve relative comparison between different retrievers. Therefore, we propose a subset sampling method to generate Pseudo GT. 

\begin{figure}[h!]
\centering 
\includegraphics[width=0.43\textwidth]{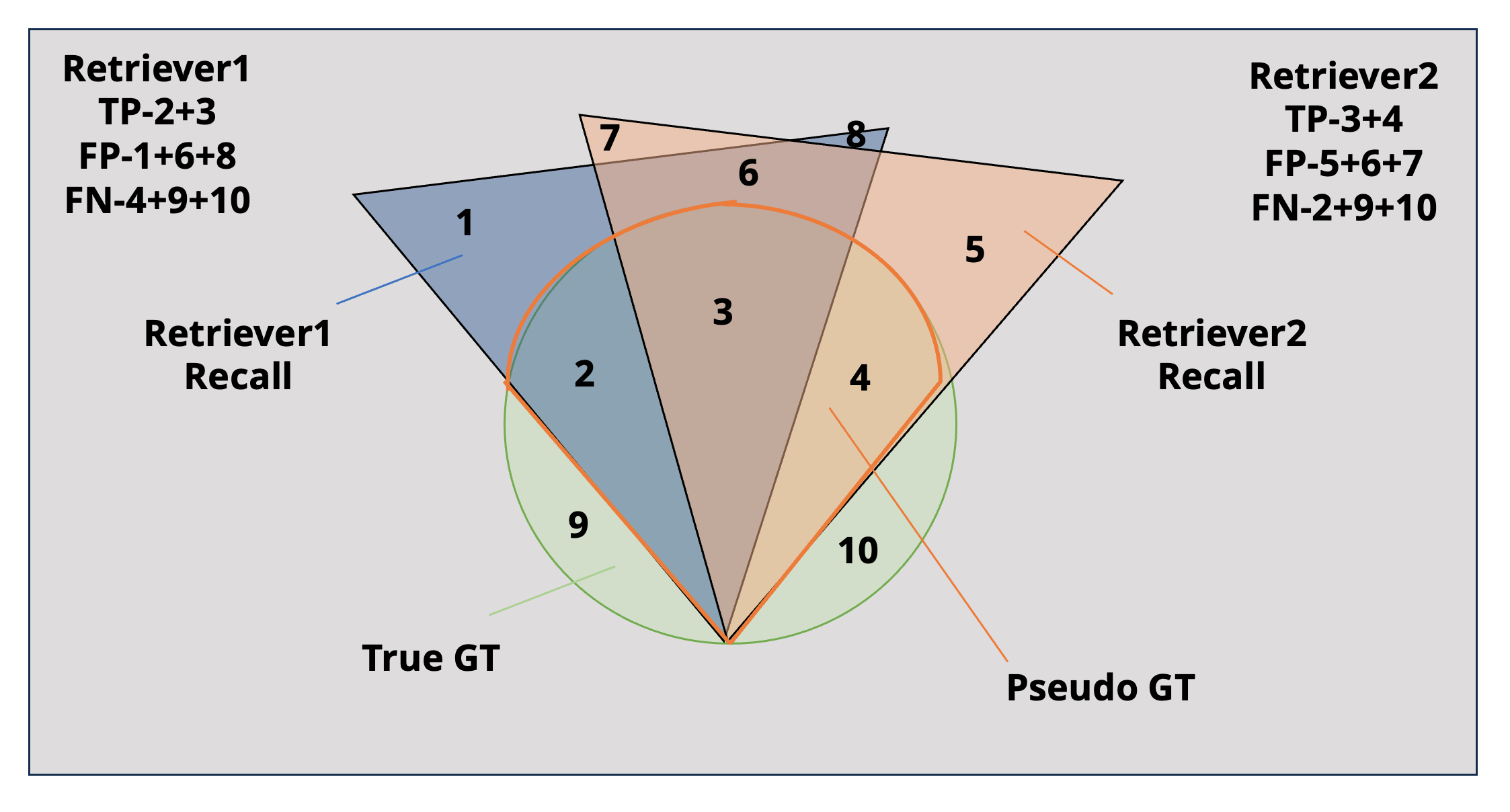}
\caption{With our subset sampling method, we can get relative comparison between different retrievers.}
\label{fig:subset_sample.png}
\end{figure}

Using the recall results of all retrievers in comparison as a sampled subset of the knowledge base, we can ensure that the number of positive samples predicted by each retriever remains unchanged, while negative samples are reduced to the same degree. As Figure 1 shows, for retrievers 1 and 2 in comparison, we aggregate all chunks recalled by both retrievers as a sampled subset of the knowledge base, thereby ensuring that all positive samples (i.e., relevant chunks recalled) predicted by both retrievers appear in the sampled subset, while their predicted negative samples (i.e., relevant chunks not recalled) are reduced to the same extent, thus achieving relative comparison between different retrievers at relatively low cost.

\subsection{Evaluation Metrics}

\textbf{Core Challenge.} 
In RAG systems, segmentation strategies create inherent trade-offs where larger chunks yield higher recall but lower precision, making it challenging to select appropriate evaluation metrics for robust retriever comparison. 

The F-beta score addresses this by introducing adjustable weights $\beta$ to balance precision and recall importance\citep{goutte2005probabilistic}:

\begin{equation}
F_{\beta} = \frac{(1 + \beta^2) \cdot \text{precision} \cdot \text{recall}}{\beta^2 \cdot \text{precision} + \text{recall}}
\end{equation}

but determining a universal $\beta$ value remains difficult for business-specific RAG systems due to their highly specialized characteristics.

\textbf{We use: PR-AUC}. PR-AUC(Precision-Recall Area Under Curve) evaluates model performance via the area under the precision-recall curve, which plots precision (y-axis) against recall (x-axis) across classification thresholds\citep{davis2006relationship}.
Unlike ROC-AUC, PR-AUC is better for imbalanced data—common in retriever evaluation, where relevant chunks (positives) are far fewer than irrelevant ones (negatives). ROC’s False Positive Rate is insensitive to such imbalance, masking retriever differences, while PR curves focus on positive identification, better reflecting retrieval ability.

Crucially, PR-AUC avoids preset precision-recall weights, using curves to show performance across thresholds for easy retriever comparison: a curve mostly above another indicates superior performance. This avoids F-beta’s $\beta$-determination issues and F1’s rigid weights, offering a robust, objective retriever evaluation standard. 

\subsection{Cross-Retriever Evaluation}

When conducting comparisons between retrievers with different segmentation strategies, the varying lengths of chunks introduce computational errors. 

\begin{figure}[h!]
\centering 
\includegraphics[width=0.43\textwidth]{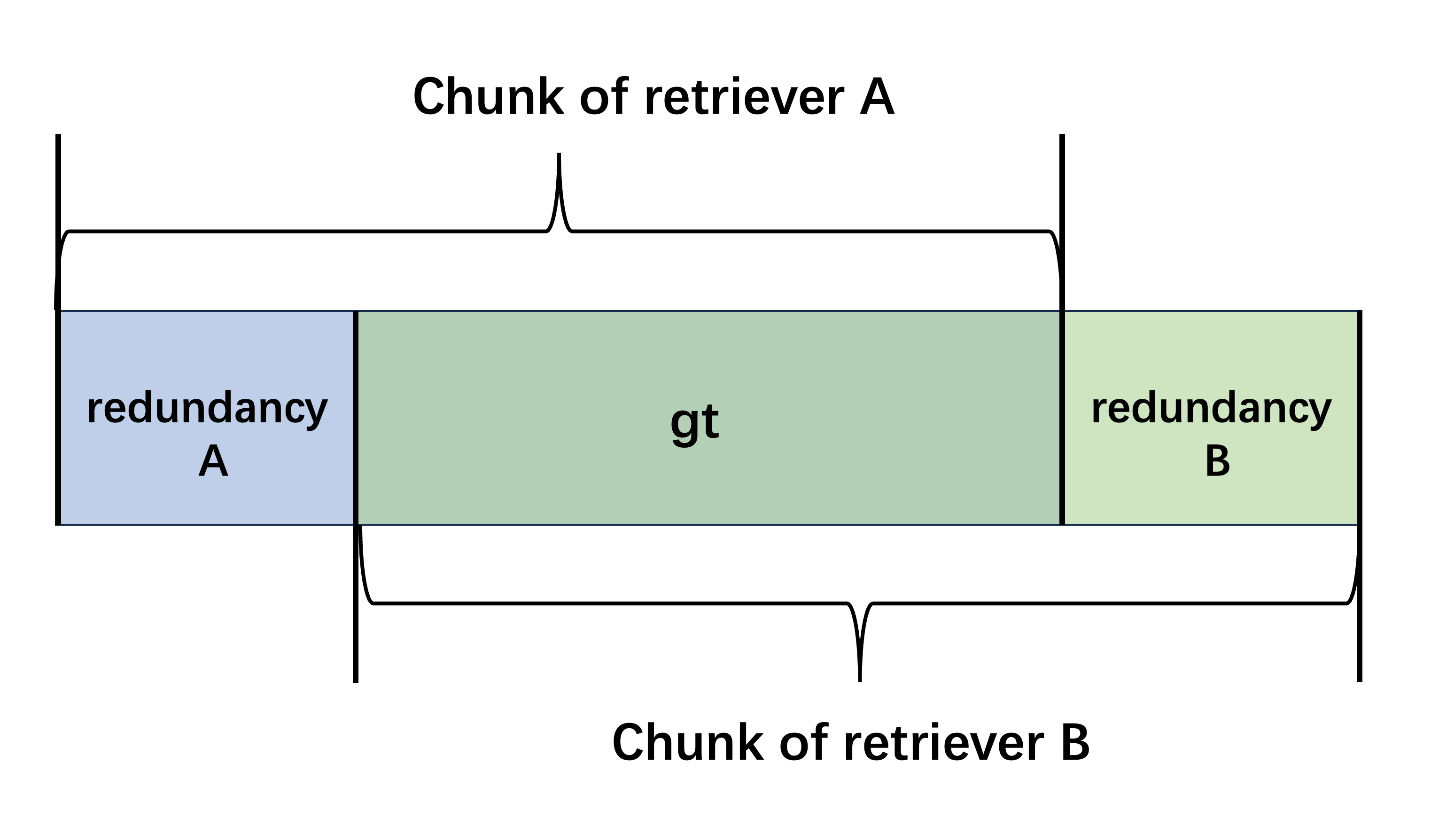}
\caption{The challenge of comparison between retrievers with different segmentation strategy.}
\label{fig:fact_extract.png}
\end{figure}

For instance, consider a comparison between retriever A using segmentation strategy A and retriever B using segmentation strategy B in Figure 2. For a specific query recalling a particular overlapping gt, retriever A recalls the segment "$redundancy A \Vert gt$", while retriever B recalls the segment "$gt \Vert redundancy B$". Both segments would be classified as relevant by the LLM and thus added to the GT subset, as both contain valid information for answering the question. However, during Pseudo GT generation, "$redundancy A \Vert gt$" and "$gt \Vert redundancy B$" would be counted as two distinct gt chunks, leading to the erroneous conclusion that retriever A failed to recall "$gt \Vert redundancy B$" and retriever B failed to recall "$redundancy A \Vert gt$". Such computational results are unreasonable, as both retriever A and retriever B have effectively recalled the essential gt.

Therefore, this paper proposes an LLM-based minimal retrieval facts generation method, utilizing LLMs to extract minimal chunks from all chunks classified as relevant. The method requires LLMs to semantically segment recalled chunks, removing redundant text to generate accurate and minimal relevant text blocks as gt segments.

For example, in the case above, the LLM would segment the recalled "$redundancy A \Vert gt$", identifying "redundancy A" as irrelevant redundant text and extracting the minimal relevant text block gt. Similarly as recalled "$gt \Vert redundancy B$", thereby ensuring that both retriever A and retriever B successfully recall gt during precision and recall calculations, achieving universal relative evaluation across retrievers.

\section{Cases}
Given the current state of RAG system applications in industry, this paper selects two representative RAG application within rednote: the knowledge-based QA application represented by "Linghang Shu" and the generative retrieval application in travel scenarios. We employ the proposed SEARA method to automatically obtain optimal retrievers for both application.

\subsection{Linghang Shu — rednote Employee Self-Service Consultation Assistant}

Linghang Shu is a rednote internal employee self-service consultation assistant based on RAG systems. It serves all employees by retrieving the enterprise's knowledge base and providing intelligent QA, boosting workplace issue processing efficiency. Primarily, it answers queries via knowledge base retrieval, with final responses generated by LLMs, accurate retrieval of query-relevant enterprise information is critical to its performance. As a deployed AI app at rednote, it maintains a database of internal documents as sources., and it's user demands focus on consultation and QA, making it a classic RAG application in knowledge-based Q\&A scenarios.

\subsubsection{Retriever Construction Strategies}
\
\newline
\indent Based on the collected enterprise documents of Linghang Shu, this paper constructs the following strategy space. Any sampling within this strategy space can construct a retriever, and our objective is to automatically obtain the optimal retriever construction strategy within this strategy space. The retriever strategy space for Linghang Shu includes three dimensions: segmentation strategy, retrieval strategy, and embedding model.

\begin{table*}
  \centering
  \begin{tabular}{lll}
    \hline
    \textbf{Strategy Dimension} & \textbf{Sub-strategy}\\
    \hline
    Segmentation Strategy & Original segmentation strategy (original) \\
    Segmentation Strategy & No merging and no secondary segmentation (nmns) \\
    Segmentation Strategy & No merging but secondary segmentation (nms) \\
    \hline
    Retrieval Strategy & dense\\
    Retrieval Strategy & hybrid\\
    \hline
    Embedding Model & tao8k, conan, xiaobu, zpoint \\
    \hline
  \end{tabular}
  \caption{Retriever Strategies}
  \label{tab:retriever-strategies}
\end{table*}

Based on combinations of sub-strategies across these three dimensions, this paper constructs 24 retrievers in total in Table 1. The detail of sub-strategies has shown in appendix B.

Our evaluation queries based on 266 real user queries obtained from online service. We employ the SEARA method to obtain the optimal retriever.

\subsubsection{Evaluation Results}

Our comprehensive evaluation of different embedding models reveals that model selection significantly impacts retrieval performance, with the conan embedding model demonstrating substantial advantages across most strategy combinations (Figure 3). The effectiveness of embedding models varies depending on the segmentation and retrieval strategies employed, with conan showing particularly pronounced advantages when combined with sparse retrieval methods. 

\begin{figure*}[h!]
\centering 
\includegraphics[width=0.8\textwidth]{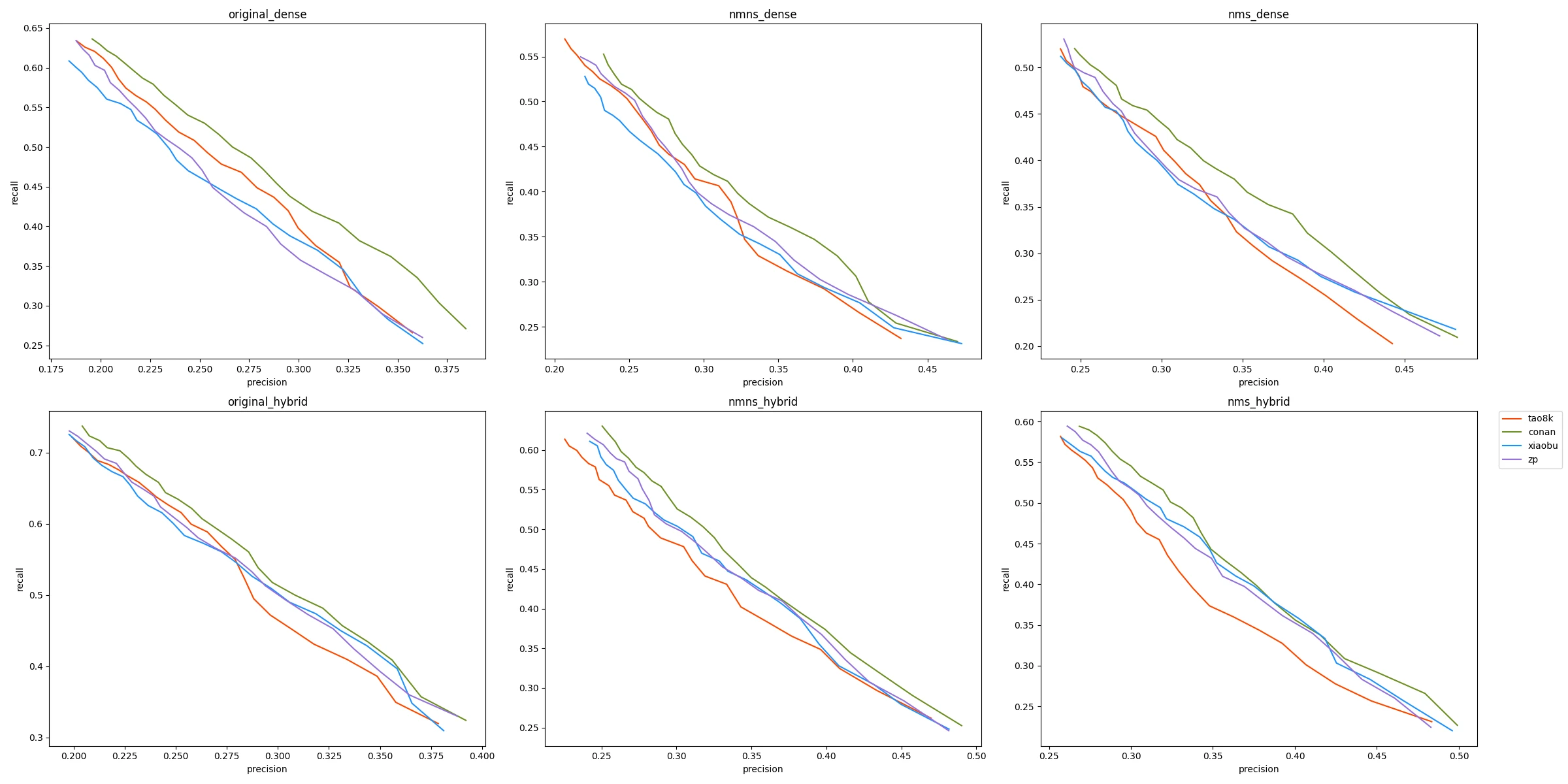}
\caption{Evaluation Result of Embedding Model.}
\label{fig:c1-3.png}
\end{figure*}

Based on the complete experimental results, the optimal retriever configuration combines the nms segmentation strategy (no merging but secondary segmentation), the conan embedding model, and hybrid retrieval incorporating sparse retrieval methods. This configuration achieves superior retrieval performance in both precision and recall metrics, leading to the optimization of Linghang Shu's RAG system for knowledge-based QA applications. Detailed results for strategy evaluations are provided in the Appendix D.

\subsection{Retrieval-based Travel Assistant}

Travel content ecosystem notes are a key component of the rednote community. Leveraging high-quality travel notes from the community and combining LLMs for generative retrieval enables the production of high-quality, accurate, and authentic personalized travel guides tailored to users' individualized needs in scenarios such as travel planning, destination exploration, and guide searching, thereby enhancing the efficiency and experience of users' travel decision-making. As a form of generalized RAG application, generative retrieval in this context relies heavily on understanding user needs and retrieving highly relevant, authentic travel notes; thus, evaluating various retriever configurations against real user queries in travel scenarios to identify optimal retrievers can significantly improve the performance of such applications.

\subsubsection{Retriever Construction Strategies}
\
\newline
\indent Generative retrieval relies on rednote community note resources. Considering the textual characteristics of notes, all rednote community note content can be viewed as a pre-constructed knowledge base, with one note corresponding to one chunk. Therefore, the retriever's objective is to recall several note contents (chunks) most relevant to a specific user query. Based on the characteristics of generative retrieval, this paper constructs different retrievers through chain combinations of query rewriters, filters, and rerankers. The retriever strategy space in this scenario includes three core dimensions: query rewriter, filter, and reranker.

\begin{table*}
  \centering
  \setlength{\tabcolsep}{4pt} 
  \renewcommand{\arraystretch}{0.8} 
  \begin{tabular}{lll}
    \hline
    \textbf{Stage} & \textbf{Sub-strategy} & \textbf{Notes} \\
    \hline
    Stage 1 & Quality filtering & Filter low quality notes \\
    Stage 1 & POI filtering & Filter bottom 25\% POI notes \\
    Stage 1 & Length filtering & Filter notes with length < 50 \\
    Stage 1 & Engagement filtering & Filter notes with interactions < 25 \\
    \hline
    Stage 2 & Qwen3-Reranker & Calculate chunk relevance scores \\
    Stage 2 & LLM-Reranker & Direct reranking with Qwen3-8B-Instruct \\
    \hline
    Stage 3 & Query rewriter & Use dots.llm1 model for query rewriting \\
    Stage 3 & Filter & Apply optimal filter from Stage 1 \\
    Stage 3 & Reranker & Apply optimal reranker from Stage 2 \\
    \hline
  \end{tabular}
  \caption{Retriever Strategies of Three Stage}
  \label{tab:retriever-strategies-three-stage}
\end{table*}

Based on combinations of sub-strategies across these three dimensions, theoretically 96 different retrievers can be constructed. Considering evaluation time costs and practicality, this paper adopts a staged evaluation strategy to optimize retriever construction in Table 2. The detail of sub-strategies has shown in Appendix C. Our evaluation queries are based on 300 real user queries collected in travel scenarios, and we use travel note content from the rednote platform to construct the knowledge base.

\subsubsection{Evaluation Results}

Our comprehensive evaluation of the retrieval pipeline examined the complex interactions between query rewriter, filter, and reranker components across 8 different combination strategies. The results reveal that these components exhibit intricate interdependencies, where the effectiveness of each component depends heavily on the states of others, in Figure 4, 5, 6. Query rewriter tends to decrease retrieval relevance by introducing results from derived queries, showing improvement only when both filter and reranker are simultaneously enabled. The reranker consistently improves performance in most scenarios. Filter components demonstrate the ability to enhance recall, but can substantially reduce precision when reranker is enabled and query rewriter is disabled.

\begin{figure*}[h!]
\centering 
\includegraphics[width=0.9\textwidth]{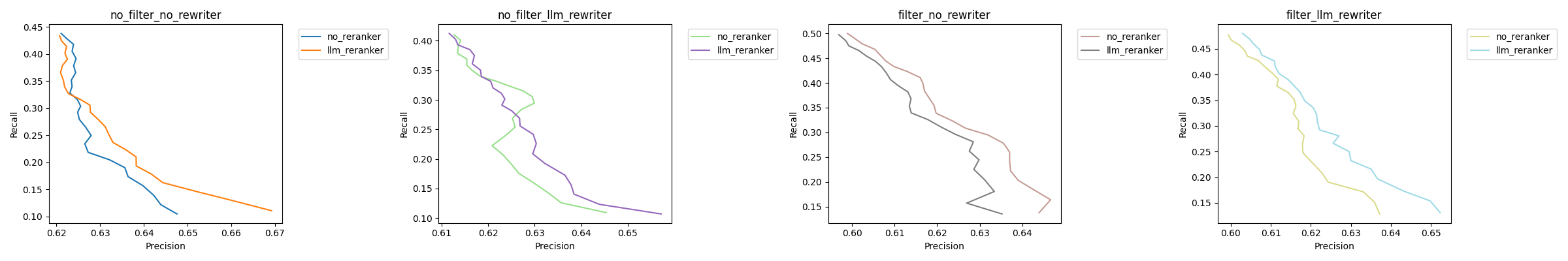}
\caption{Evaluation Result of Using Rewriter.}
\label{fig:c1-1.png}
\end{figure*}

\begin{figure*}[h!]
\centering 
\includegraphics[width=0.9\textwidth]{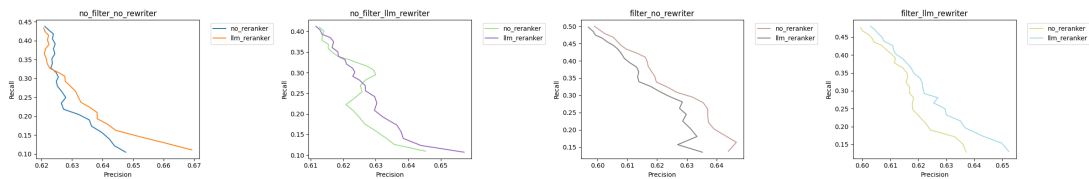}
\caption{Evaluation Result of Using Reranker.}
\label{fig:c1-1.png}
\end{figure*}

\begin{figure*}[h!]
\centering 
\includegraphics[width=0.9\textwidth]{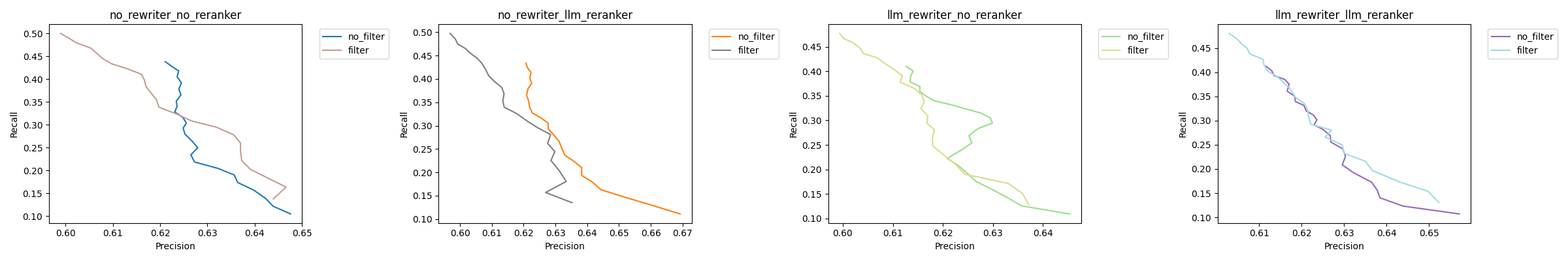}
\caption{Evaluation Result of Using Filter.}
\label{fig:c1-1.png}
\end{figure*}

Based on our evaluation, two optimal retrievals emerge for travel assistant: (1) simultaneously enabling query rewriter + filter + reranker for precision-prioritized applications, and (2) enabling only filter for recall-prioritized applications. The optimal filter combines POI filter, length filter, and engagement filter while disabling quality filter. The optimal reranker is LLM reranker. These findings highlight the critical importance of considering component interactions rather than optimizing individual components in isolation when designing RAG retrieval pipelines. Detailed evaluation results for individual filter components, reranker comparisons, and component interaction analyses are provided in the Appendix E.

\section{Related work}

\textbf{Retrieval-Augmented Generation (RAG) Evaluation}
RAG integrates LLMs through a retriever-generator architecture~\citep{lewis2020retrieval, gao2023retrieval}, reducing hallucinations and improving response accuracy~\citep{huang2025survey, yu2024evaluation}. RAG evaluation is complex, requiring assessment of each components and overall performance. Computing retrieval gt incurs high costs~\citep{tang2024multihop}, generators require factual correctness and relevance evaluation~\citep{zheng2023judging}, and component interactions necessitate end-to-end assessment~\citep{es2024ragas}. Downstream task complexity and production factors like latency must also be considered~\citep{zhang2022survey, chen2024benchmarking}.Yu et al. proposed a three-dimensional framework contains retriever, generator and overall system~\citep{yu2024evaluation}, but specific methods for efficient retriever evaluation without gt remain gaps, which this paper addresses.

\textbf{RAG Evaluation Benchmarks}
Researchers have constructed targeted benchmarks drawing from established datasets like KILT~\citep{petroni2020kilt} and SuperGLUE~\citep{wang2019superglue}. Chen et al. developed RGB evaluating LLMs' four basic RAG capabilities~\citep{chen2024benchmarking}. Lyu et al. created CRUD-RAG covering four application scenarios~\citep{lyu2025crud}. Tang et al. constructed MultiHop-RAG for multi-hop queries~\citep{tang2024multihop}.
However, benchmark-only evaluation cannot address two issues: real user queries contain complex non-factual or ambiguous questions challenging standard answer construction, and scenario-specific requirements prevent universal benchmark.

\textbf{RAG Automated Evaluation Methods}
LLM-based frameworks address standard answer construction difficulties. Ragas generates questions and answers from documents for evaluation~\citep{es2024ragas}. ARES fine-tunes models for domain-specific evaluation~\citep{saad2023ares}. RAGEval uses pattern-based processes for high-quality sample generation~\citep{zhu2024rageval}.
However, synthetic data is often overly standardized, failing to reflect real-world query complexity and affecting evaluation effectiveness.

\textbf{Retriever Evaluation}
Existing RAG evaluation frameworks adopt simplified retriever evaluation strategies. Ragas measures context relevance using LLM judgment but ignores recall effectiveness~\citep{es2024ragas}. CRUD-RAG uses MRR considering only first retrieval gt, insufficient for multi-document synthesis needs~\citep{lyu2025crud}. MultiHop-RAG employs Hit Rate but fails to accurately reflect complete evidence chain collection~\citep{tang2024multihop}.
Traditional metrics like precision and recall face adaptation difficulties in RAG scenarios, Limited attempts like Xu et al.'s F1-recall application face construction difficulties and high costs~\citep{xu2024let}, with evaluation results often differing from actual performance.

\section{Conclusion}

Addressing the challenge of determining optimal retriever construction strategies for specific business scenarios in RAG systems, this paper proposes the SEARA method. Through the utilization of real user queries, this method enables fully automated retriever evaluation, thereby identifying optimal retrievers for specific business contexts, providing a data-driven solution for RAG system performance improvement. Based on this method, we automatically obtain optimal retrievers in two scenarios in rednote. Experimental results demonstrate that optimal retrievers are highly application-dependent, the automated acquisition of optimal strategy combinations tailored to diverse application environments through the SEARA method holds significant implications for the practical performance of RAG applications.
\section{Limitation}

\textbf{Trade-off between computational efficiency and evaluation accuracy}: Our proposed SEARA method faces a challenge in balancing computational cost with evaluation quality. When constructing Pseudo GT through subset sampling, incorporating a larger number of retrievers in the comparison process yields Pseudo GT that more closely approximates the true GT, thereby enhancing evaluation reliability. However, this improved accuracy comes at the expense of significantly increased computational overhead, as each additional retriever requires substantial processing time and resources. This computational burden becomes particularly pronounced in large-scale retrieval systems or when frequent evaluations are necessary. From a practical standpoint, different application scenarios may require different evaluation strategies - real-time systems prioritize computational efficiency while offline evaluation systems can afford more comprehensive comparisons. Future work should focus on developing adaptive mechanisms that can dynamically adjust the evaluation scope based on available computational resources and required accuracy levels.

\textbf{Dependence on authentic user queries and handling of unanswerable queries}: SEARA's effectiveness is inherently tied to the availability and quality of authentic user query datasets, which may not always be accessible or representative. The method assumes that user queries have retrievable answers within the knowledge base, but real-world scenarios often include unanswerable queries - those for which no relevant information exists. Currently, our framework does not distinguish between answerable and unanswerable queries, which may lead to suboptimal evaluation outcomes and wasted computational resources. This limitation stems from the challenge of automatically detecting query answerability without access to ground truth annotations. Developing robust mechanisms to identify and appropriately handle unanswerable queries represents a promising direction for future work, potentially through query difficulty estimation or retrieval confidence scoring techniques.

\bibliography{custom}

\appendix

\section{Mathematical Proof of Subset Sampling}
\label{sec:appendix}

Here we demonstrate through metric calculations that the results computed based on Pseudo GT are relatively consistent with those computed on real GT. 

Assume that on real GT, retriever 1 yields $TP_1$, $FP_1$, $TN_1$, $FN_1$, and retriever 2 yields $TP_2$, $FP_2$, $TN_2$, $FN_2$. Assume that on the Pseudo GT generated in this paper, retriever 1 yields $TP_1'$, $FP_1'$, $TN_1'$, $FN_1'$, and retriever 2 yields $TP_2'$, $FP_2'$, $TN_2'$, $FN_2'$.

Based on the Pseudo GT generation method using subset sampling proposed in this paper, all positive examples predicted by retrievers will appear in the subset, therefore:

\begin{equation}
TP_1 = TP_1'
\end{equation}
\begin{equation}
FP_1 = FP_1'
\end{equation}
\begin{equation}
TP_2 = TP_2'
\end{equation}
\begin{equation}
FP_2 = FP_2'
\end{equation}

Since samples recalled by any retriever will appear in the subset, positive samples not appearing in the subset are $FN_{res}$ (actually relevant samples not recalled by any retriever), and negative samples not appearing in the subset are $TN_{res}$ (actually irrelevant samples that were also not mistakenly recalled by any retriever). According to the subset sampling method:
\begin{equation}
FN_1' = FN_1 - FN_{res}
\end{equation}
\begin{equation}
TN_1' = TN_1 - TN_{res}
\end{equation}
\begin{equation}
FN_2' = FN_2 - FN_{res}
\end{equation}
\begin{equation}
TN_2' = TN_2 - TN_{res}
\end{equation}

Since the knowledge base is fixed, all positive samples are fixed. Let all positive samples be $P$, then:

\begin{equation}
TP_1 + FN_1 = P
\end{equation}
\begin{equation}
TP_1' + FN_1' = P - FN_{res}
\end{equation}

Assume that the precision calculated by retriever 1 on real GT and Pseudo GT are $precision_1$ and $precision_1'$, and the recall are $recall_1$ and $recall_1'$, respectively. Similarly, assume that the precision and recall calculated by retriever 2 on real GT and Pseudo GT are $precision_2$, $precision_2'$, $recall_2$, and $recall_2'$. According to the calculation principle of precision:

\begin{equation}
\begin{split}
precision_1 &= \frac{TP_1}{TP_1+FP_1} \\
&= \frac{TP_1'}{TP_1'+FP_1'} \\
&= precision_1'
\end{split}
\end{equation}

\begin{equation}
\begin{split}
precision_2 &= \frac{TP_2}{TP_2+FP_2} \\
&= \frac{TP_2'}{TP_2'+FP_2'} \\
&= precision_2'
\end{split}
\end{equation}

Therefore, the precision calculated on Pseudo GT is completely consistent with the real precision.

According to the calculation principle of recall:

\begin{equation}
\begin{split}
recall_1 &= \frac{TP_1}{TP_1 + FN_1} \\
&= \frac{TP_1'}{TP_1' + FN_1' + FN_{res}} \\
&= \frac{TP_1'}{P}
\end{split}
\end{equation}

\begin{equation}
\begin{split}
recall_1' = \frac{TP_1'}{TP_1'+FN_1'} = \frac{TP_1'}{P - FN_{res}}
\end{split}
\end{equation}

\begin{equation}
\begin{split}
recall_2 &= \frac{TP_2}{TP_2 + FN_2} \\
&= \frac{TP_2'}{TP_2' + FN_2' + FN_{res}} \\
&= \frac{TP_2'}{P}
\end{split}
\end{equation}

\begin{equation}
\begin{split}
recall_2' = \frac{TP_2'}{TP_2'+FN_2'} = \frac{TP_2'}{P - FN_{res}}
\end{split}
\end{equation}

To demonstrate that recall calculated on Pseudo GT is relatively comparable, we assume:

\begin{equation}
recall_1' > recall_2'  
\end{equation}

That is:

\begin{equation}
\frac{TP_1'}{P - FN_{res}} > \frac{TP_2'}{P - FN_{res}}
\end{equation}

\begin{equation}
TP_1' > TP_2'
\end{equation}

Then:

\begin{equation}
\frac{TP_1'}{P} > \frac{TP_2'}{P}
\end{equation}

\begin{equation}
\frac{TP_1'}{P} > \frac{TP_2'}{P}
\end{equation}

\begin{equation}
recall_1 > recall_2
\end{equation}

This demonstrates that based on the subset sampling method proposed in this paper, the recall calculated on Pseudo GT is relatively comparable to the recall on real GT. Meanwhile, assuming each retriever recalls top-K relevant chunks, and generally $K << N$, the computational cost for each retriever based on the subset sampling method is $M\times K << M\times N$, thus achieving relative evaluation of different retrievers at low cost.

\section{Detail of Linghang Shu Retriever Strategies Space}
\label{sec:appendix}

\textbf{Segmentation Strategy} 

Segmentation strategy refers to how to perform chunk segmentation on enterprise source documents. The core evaluation point of this dimension is whether to control chunk length distribution through secondary segmentation, merging, and other operations. Sub-strategies in this dimension include:

\begin{itemize}
\item \textbf{Original strategy (original)}: Considering that Linghang Shu primarily uses source documents in standardized reddoc format (rednote's internal cloud document), and to control chunk length concentration, our defined original strategy segments according to document title hierarchy, merges adjacent small chunks of the same hierarchy level, and performs secondary segmentation on overly long chunks.
\item \textbf{No merging but secondary segmentation (nms)}: Similarly segments according to document title hierarchy and performs secondary segmentation on overly long chunks, but avoids merging operations to prevent potential topic mixing issues caused by merging small chunks.
\item \textbf{No merging and no secondary segmentation (nmns)}: Only segments according to document title hierarchy without merging operations, and simultaneously avoids secondary segmentation to prevent potential topic fragmentation issues caused by secondary segmentation.
\end{itemize}

\textbf{Retrieval Strategy} 

Retrieval strategy refers to how to recall chunks from the enterprise knowledge base constructed based on segmentation strategy for given user queries. The core evaluation point of this dimension is whether to combine sparse retrieval methods to improve retrieval performance. Sub-strategies in this dimension include:

\begin{itemize}
\item \textbf{Dense retrieval (dense)}: Directly uses dense embedding models to convert chunks into semantic vectors, then recalls the top K chunks most semantically similar to the query through HNSW (Hierarchical Navigable Small Worlds) algorithm.

\item \textbf{Hybrid retrieval (hybrid)}: Uses a hybrid retrieval method combining BM25 sparse retrieval based on keyword matching to assist dense retrieval in recalling the top K chunks most semantically similar to the query.
\end{itemize}

\textbf{Embedding Model} 

Refers to which dense embedding model is used to achieve chunk-to-semantic vector conversion. This paper selects four mainstream embedding models with good Chinese retrieval performance:

\begin{itemize}

\item \textbf{Conan-embedding (conan)\footnote{https://huggingface.co/TencentBAC/Conan-embedding-v1}}: An open-source Chinese embedding model trained and released by Tencent, based on BERT architecture and trained through contrastive learning with dynamic hard negative sampling methods.

\item \textbf{Tao-8k-embedding (tao8k)\footnote{https://huggingface.co/Amu/tao-8k}}: Based on the Stella-v2 model, extending context to 8K and trained using Chinese data. 

\item \textbf{Zpoint-embedding (zpoint)\footnote{https://huggingface.co/iampanda/zpoint\_large\_embedding\_zh}}: Also based on the Stella-v2 model, using LLM-synthesized data for multi-task training in retrieval tasks.

\item \textbf{Xiaobu-embedding (xiaobu)\footnote{https://huggingface.co/lier007/xiaobu-embedding}}: Based on the GTE model with Chinese multi-task fine-tuning.

\end{itemize}

\section{Detail of Travel Assistant Retriever Strategies Space}
\label{sec:appendix}

\textbf{Query Rewriter} 

The query rewriter refers to whether semantic expansion and optimization are performed on user-input travel-related queries to improve recall effectiveness. The core evaluation point of this dimension is whether to use a query rewriter to enhance recall effectiveness and semantic matching. This paper uses the dots.llm1 \footnote{https://huggingface.co/rednote-hilab/dots.llm1.inst} model as the query rewriter. Sub-strategies in this dimension include:

\begin{itemize}
\item \textbf{No query rewriter (no\_rewriter)}: Directly uses the user's original query for retrieval, recalling Top50 results while maintaining the originality of query intent.

\item \textbf{Using query rewriter (llm\_rewriter)}: Uses the dots.llm1 model to perform semantic expansion on the user's original query, generating 4 additional queries, with each query recalling Top10 results.

\end{itemize}

\textbf{Filter} 

The filter refers to whether to add specific conditions when recalling candidate notes. The core evaluation point of this dimension is whether to improve the relevance and quality of retrieval results through multi-level filter. This dimension includes switch combinations of four sub-dimensions:

\begin{itemize}
\item \textbf{Quality filter (quality)}: Filters low-quality content based on note text quality and image quality.

\item \textbf{Length filter (length)}: Filters content lacking informativeness based on note text length.

\item \textbf{POI filter (POI)}: Filters notes that do not include query-related locations based on the number of geographical location information mentioned in notes.

\item \textbf{Engagement filter (engage)}: Filters content lacking user recognition based on user interaction behaviors (including likes, replies, forwards).

\end{itemize}

\textbf{Reranker} 

The reranker refers to reranking candidate notes after initial retrieval and filter to prioritize relevant information in recall results. The core evaluation point of this dimension is selecting appropriate reranking models to optimize the final retrieval result order. Sub-strategies in this dimension include:

\begin{itemize}

\item \textbf{Qwen3-Reranker reranking (qwen\_reranker)}: Uses the Qwen3-Reranker-0.6B model to calculate relevance scores between queries and note content for reranking.

\item \textbf{Qwen3-8B few-shot reranking (llm\_reranker)}: Uses Qwen3-8B LLM for content reranking through few-shot in-context learning..

\item \textbf{No reranker (no\_reranker)}: Does not perform reranking, directly using the original ranking results from the retriever.

\end{itemize}

\section{Evaluation Result of Linghang Shu Q\&A Bot}
\label{sec:appendix}

\textbf{(1) Segmentation Strategy.} 

We first compare the effects of three sub-strategies for segmentation strategy, fitting other strategy selections, with results shown in the Figure 7.

\begin{figure*}[h!]
\centering 
\includegraphics[width=1\textwidth]{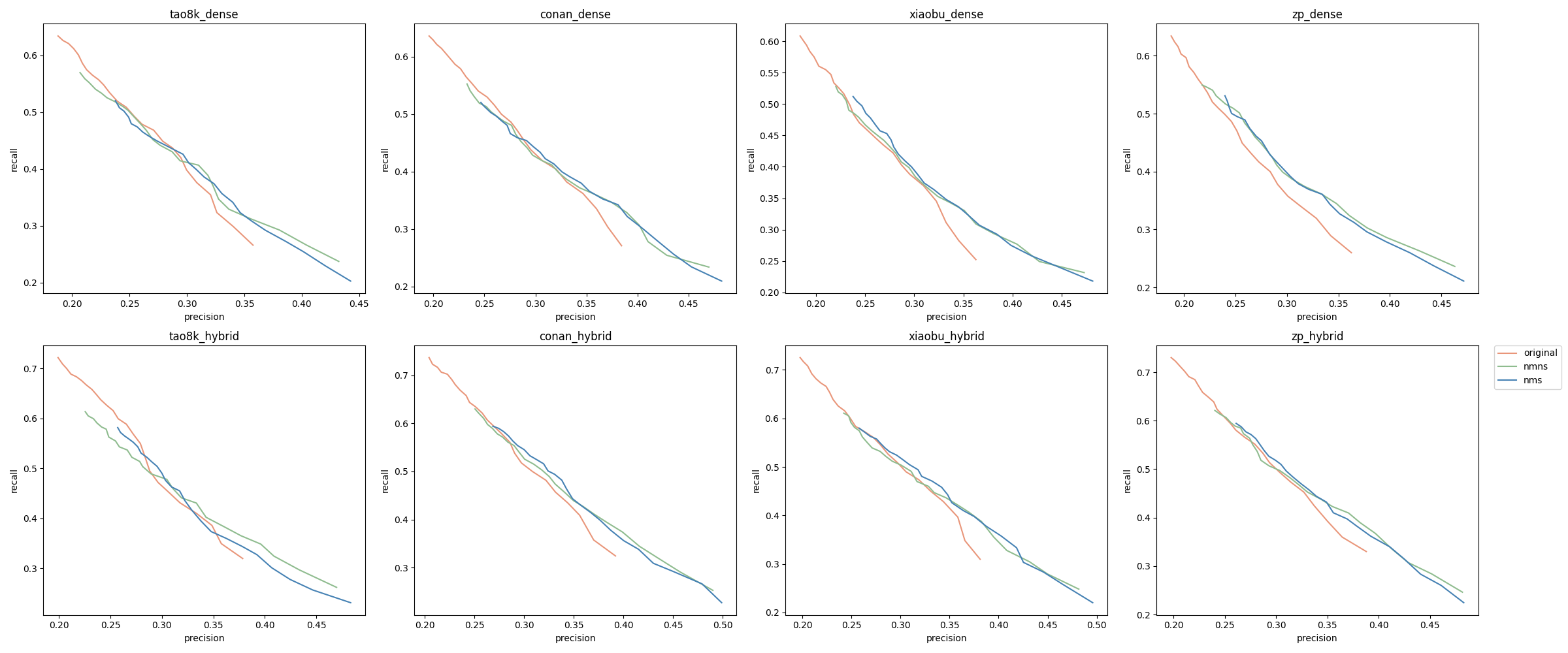}
\caption{Evaluation Result of Segmentation Strategy.}
\label{fig:c1-1.png}
\end{figure*}

It can be observed that different segmentation strategies have minimal impact on retriever retrieval effectiveness. The original segmentation strategy, which performs small text chunk merging and secondary segmentation of overly long text chunks, achieves higher recall due to its ability to ensure concentrated text chunk lengths. The nms strategy (no small text chunk merging) and nmns strategy (no merging of small text chunks and no segmentation of overly long text chunks) achieve higher precision by better avoiding semantic fragmentation. Overall, different segmentation strategies show similar retrieval performance.

\textbf{(2) Retrieval Strategy.} 

We compare two sub-strategies for retrieval strategy, fitting other strategy selections, results are shown in the Figure 8.

\begin{figure*}[h!]
\centering 
\includegraphics[width=1\textwidth]{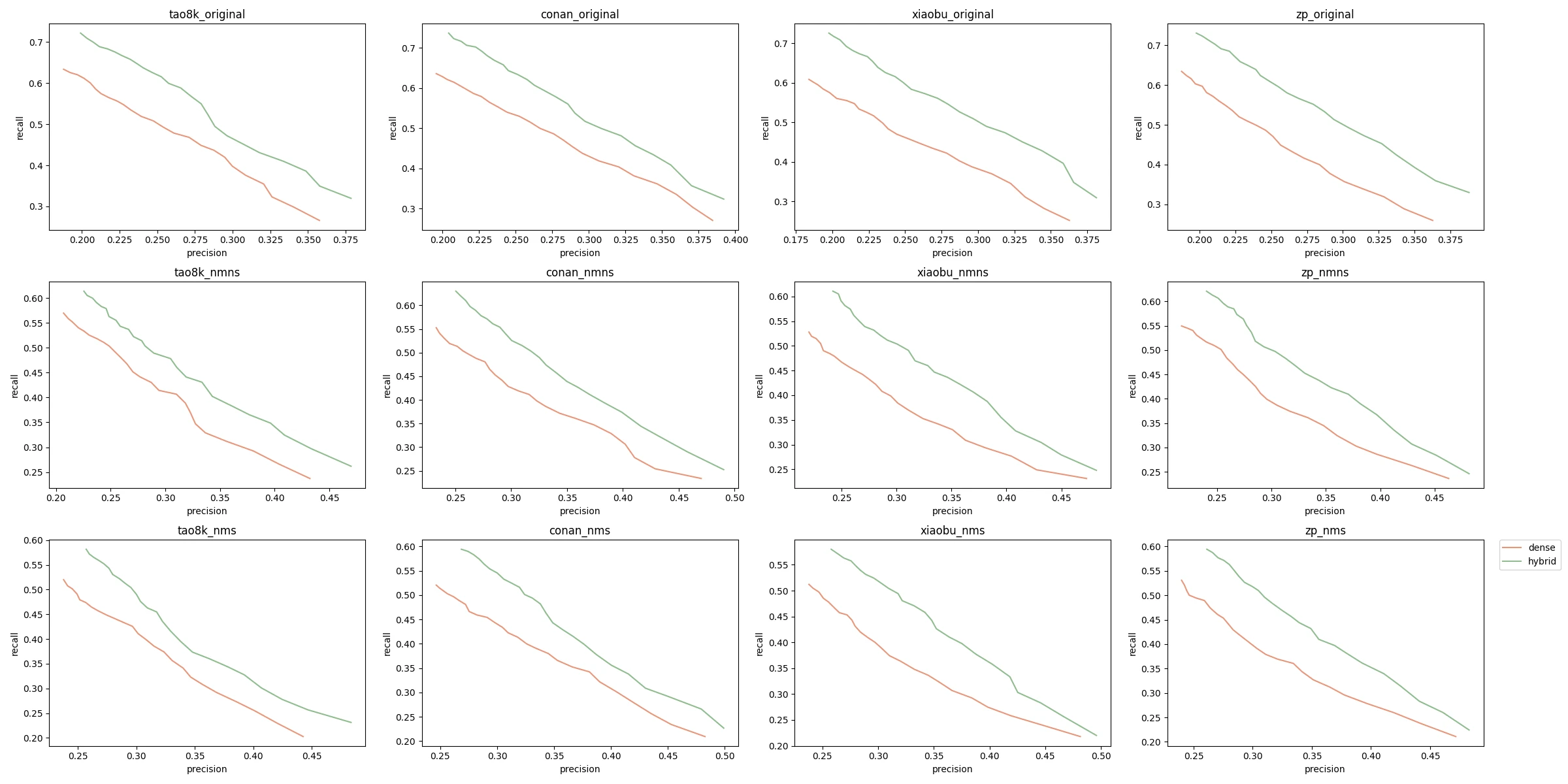}
\caption{Evaluation Result of Retrieval Strategy.}
\label{fig:c1-2.png}
\end{figure*}

It can be observed that across all strategy combinations, hybrid retrieval strategy demonstrates significant advantages, achieving higher performance in both precision and recall. This indicates that incorporating sparse retrieval methods can substantially enhance retriever retrieval performance.

\textbf{(3) Embedding Model.} 

We compare four different embedding models, fitting other strategy selections, results are shown in the Figure 9.

\begin{figure*}[h!]
\centering 
\includegraphics[width=1\textwidth]{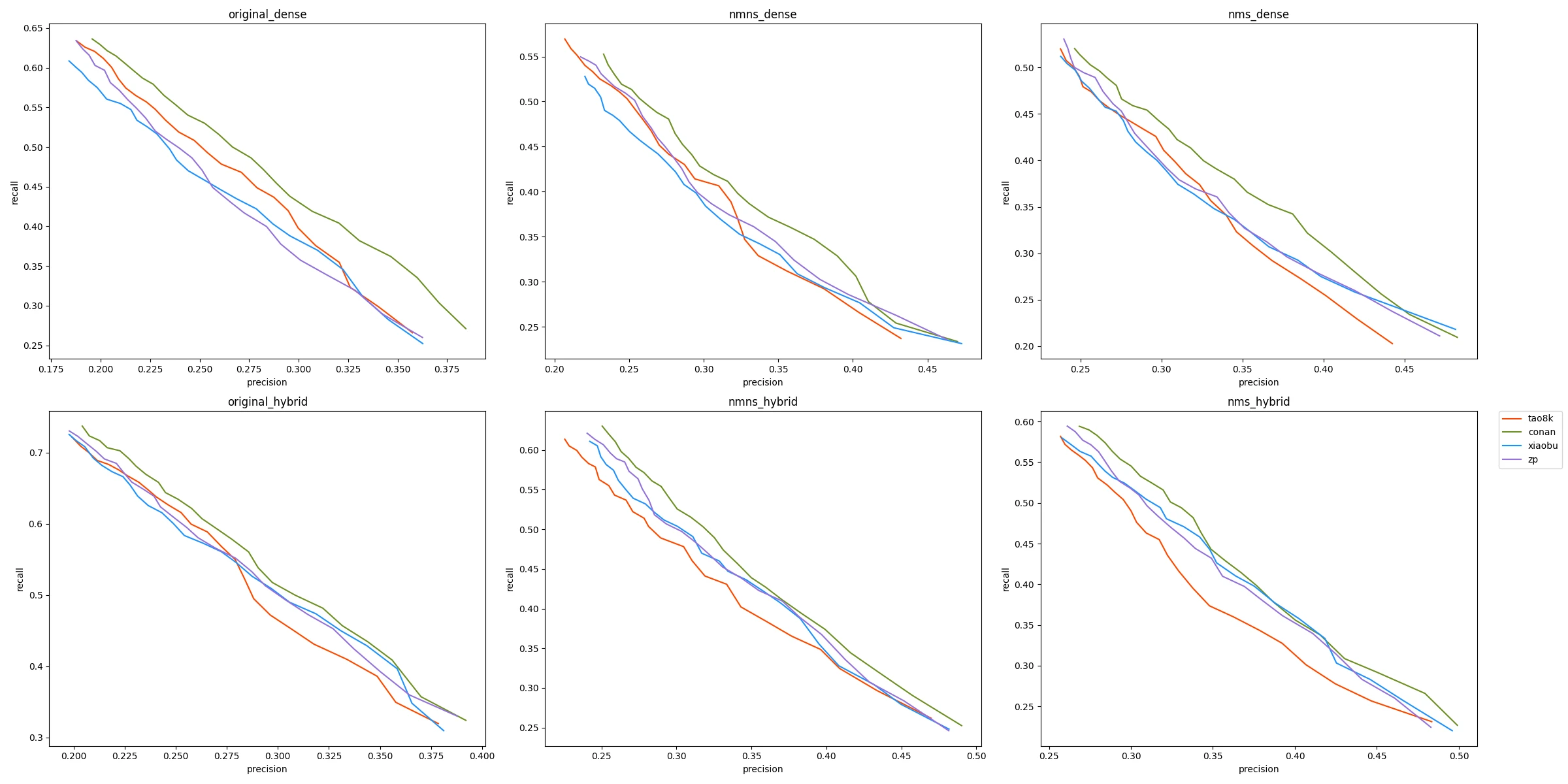}
\caption{Evaluation Result of Embedding Model.}
\label{fig:c1-3.png}
\end{figure*}

It can be observed that the effectiveness of different embedding models correlates with the segmentation and retrieval strategies used, with embedding models showing different performance advantages under different strategy combinations. However, under most strategy combinations, the conan embedding model shows significant advantages, with more pronounced advantages when the retrieval strategy is sparse retrieval. The performance differences between different embedding models significantly diminish under hybrid retrieval, indicating that the conan embedding model is most suitable for Linghang Shu's application scenario.

\section{Evaluation Result of Travel Assistant}
\label{sec:appendix}

\textbf{(1) Filter Evaluation.} 

We first compare combinations of four filter dimensions (16 strategies total) to obtain the optimal filter:

\textbf{Quality Filter}: Results of Figure 10 show that enabling quality filter produces negative effects on retrieval results in almost all scenarios.

\begin{figure*}[h!]
\centering 
\includegraphics[width=1\textwidth]{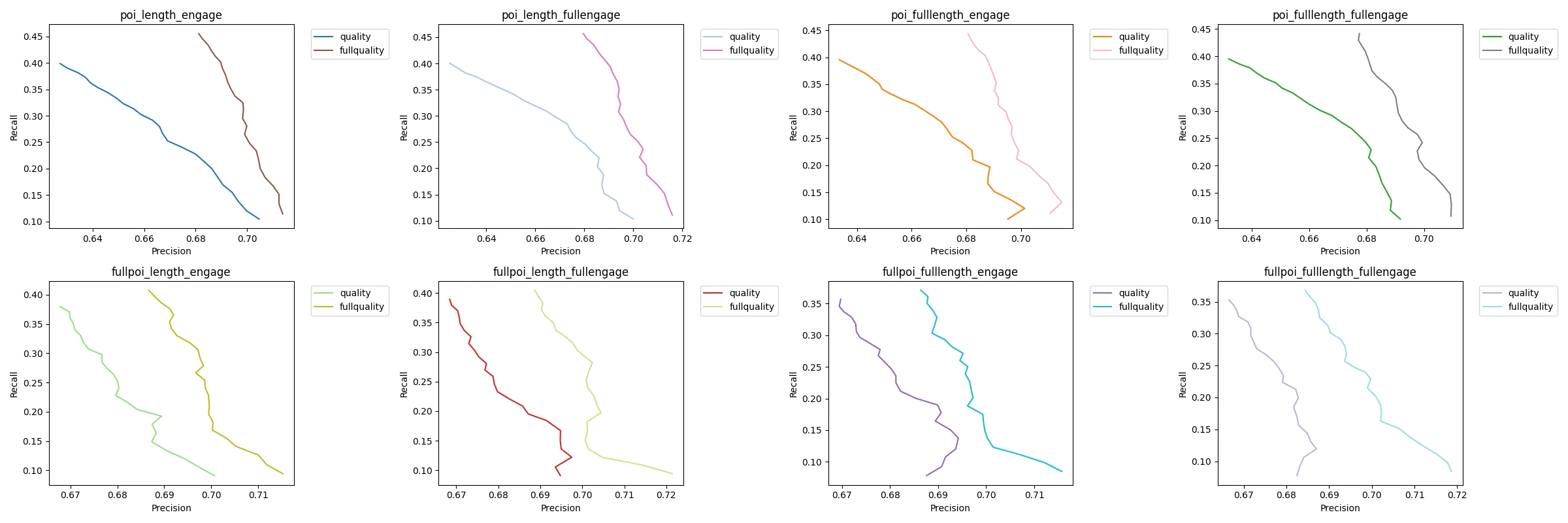}
\caption{Evaluation Result of Quality Filter.}
\label{fig:c1-1.png}
\end{figure*}

\textbf{POI Filter}: According to experimental results, enabling POI filter achieves higher recall in most scenarios but slightly lower precision. Because of the length of paper, we didn't show this figure. Overall, enabling POI filter shows better performance.

\textbf{Length Filter}: As shown in Figure 11, enabling length filter performs better in most situations.

\begin{figure*}[h!]
\centering 
\includegraphics[width=1\textwidth]{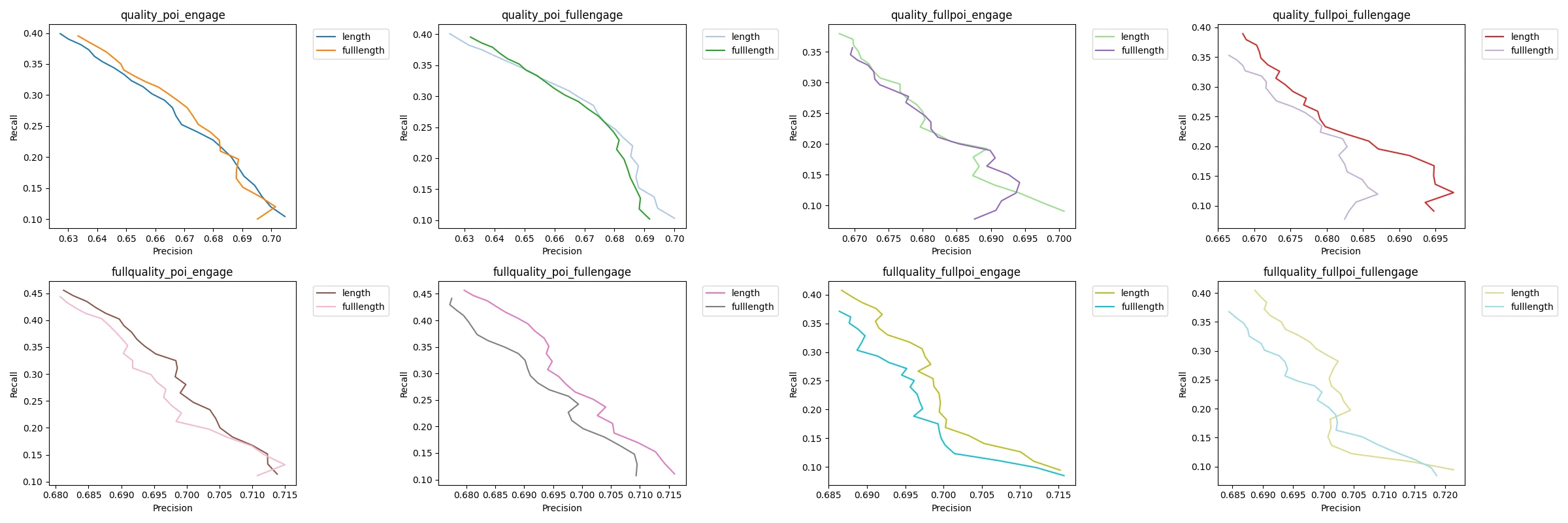}
\caption{Evaluation Result of Length Filter.}
\label{fig:c1-1.png}
\end{figure*}

\textbf{Engagement Filter}: The impact of engagement filter is complex showned in Figure 12, with curves showing poorer performance in all scenarios whether enabled or not. However, combined with the above findings, when POI filter and length filter are simultaneously enabled, enabling engagement filter achieves better final retrieval effectiveness.

\begin{figure*}[h!]
\centering 
\includegraphics[width=1\textwidth]{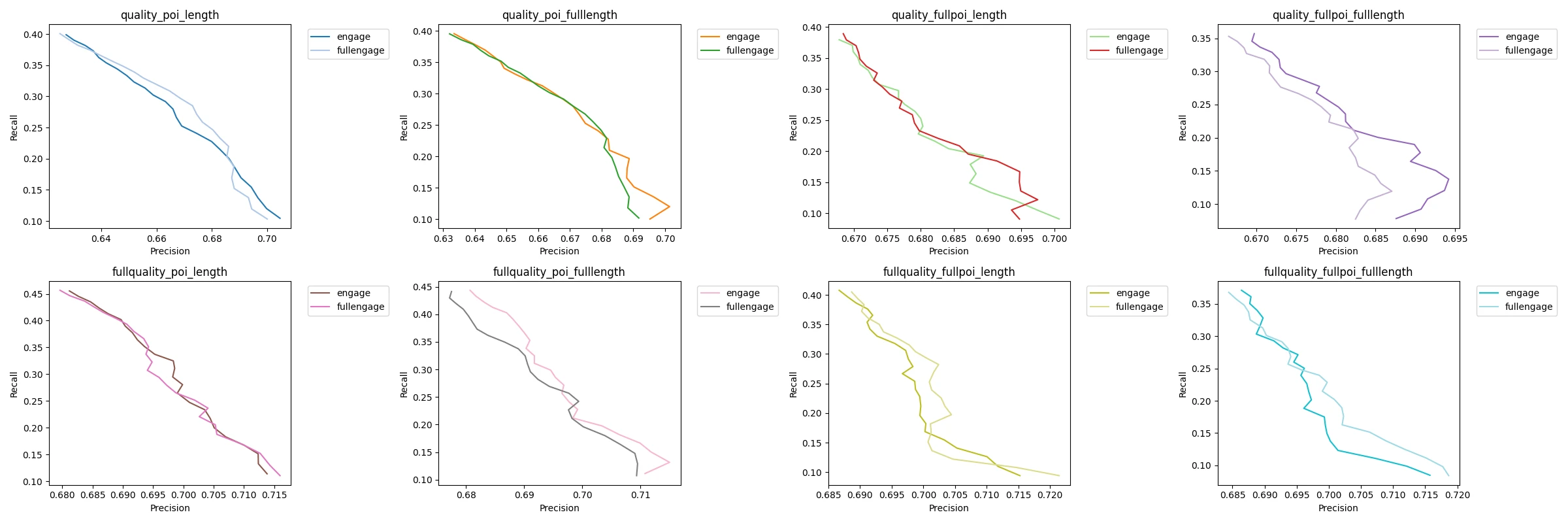}
\caption{Evaluation Result of Engagement Filter.}
\label{fig:c1-1.png}
\end{figure*}

In summary, for travel scenarios, the optimal filter enables POI filter, length filter, and engagement filter while disabling quality filter.

\textbf{(2) Reranker Evaluation} 

For two types of rerankers, we conduct comprehensive comparison using our method (filters are not used during experiments) in Figure 13.

\begin{figure*}[h!]
\centering 
\includegraphics[width=1\textwidth]{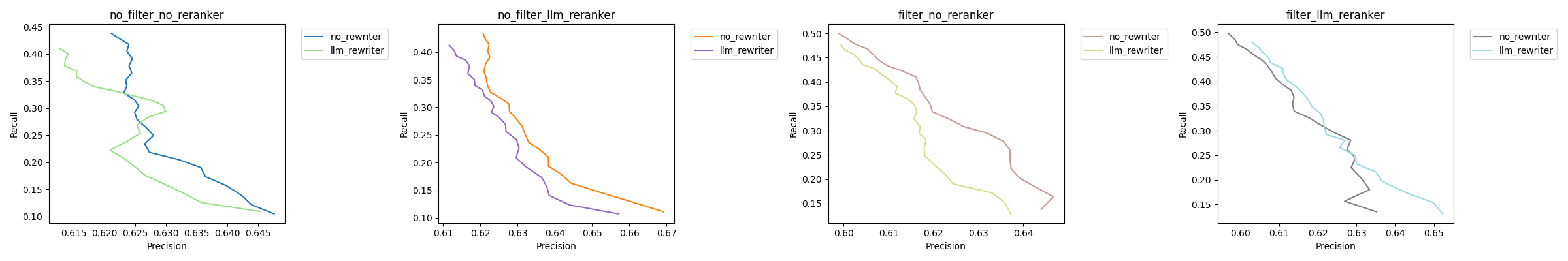}
\caption{Evaluation Result of Reranker}
\label{fig:c1-2.png}
\end{figure*}

Comparison shows that rerankers have minimal impact on results, but using LLM-Reranker still brings some degree of performance improvement. Additionally, for one query, Qwen-Reranker requires 50 calculations to complete reranking (needs to calculate relevance between each chunk and query), while LLM-Reranker only requires 4 calculations to complete reranking (uses sliding window to rerank 50 chunks with window size 20 and overlap of 10 chunks), making LLM-Reranker more cost-effective. Therefore, in the current scenario, LLM-Reranker is the optimal reranker.

\end{document}